\documentclass[nofootinbib,aps,preprint,superscriptaddress]{revtex4}
\usepackage{amsmath} 
\usepackage{citesort}
\usepackage{epsfig}

\newcommand{\beqa}{\begin{eqnarray}}
\newcommand{\eeqa}{\end{eqnarray}}
\newcommand{\beq}{\begin{equation}}
\newcommand{\eeq}{\end{equation}}
\newcommand{\bal}{\begin{align}}
\newcommand{\eal}{\end{align}}
\renewcommand{\Re}{{\rm Re\;}}

\def\gsim{\ \rlap{\raise 3pt \hbox{$>$}}{\lower 3pt \hbox{$\sim$}}\ }
\def\lsim{\ \rlap{\raise 3pt \hbox{$<$}}{\lower 3pt \hbox{$\sim$}}\ }

\begin{document}

\preprint{\vbox{
\hbox{}
\hbox{hep-ph/0505270}
\hbox{May 2005}
}}

\vspace*{48pt}
\title{The Effect of $D-\bar D$ Mixing  
on the Measurement of $\gamma$ in $B\to D K$ Decays}

\author{Yuval Grossman} 
\affiliation{Department of Physics,     
Technion--Israel Institute of Technology,\\ 
Technion City, 32000 Haifa, Israel\vspace*{3pt}}
\affiliation{Physics Department, Boston University, 
Boston, MA 02215\vspace*{3pt}}
\affiliation{Jefferson Laboratory of Physics, Harvard University,
  Cambridge, MA 02138\vspace*{3pt}}    
\author{Abner Soffer}
\affiliation{Department of Physics, Colorado State University, 
Fort Collins, CO 80523\vspace*{3pt}}
\author{Jure Zupan}
\affiliation{Department of Physics, Carnegie Mellon University,
Pittsburgh, PA 15213\vspace*{3pt}
}
\affiliation{J.~Stefan Institute, Jamova 39, P.O. Box 3000,1001
Ljubljana, Slovenia\vspace*{18pt}}

\begin{abstract} \vspace*{18pt}
$D-\bar D$ mixing is the source of the largest theoretical uncertainty
in the extraction of $\gamma$ from $B\to D K$ decays.  In the
Standard Model, the mixing can have a rate close to its current
experimental upper bound and is CP-conserving to an excellent
approximation.  We show that neglecting CP-conserving $D-\bar D$
mixing leads to an error in the determination of $\gamma$ only at
second order in the small parameters, $\Delta m_D/\Gamma_D$ and
$\Delta\Gamma_D/\Gamma_D$, and is therefore very small and can be
safely neglected.
\end{abstract}

\maketitle

The most precise determination of the standard model CKM phase
$\gamma$ will, in the long run, be provided by methods based on the
interference between $b\to c \bar{u} s$ and $b\to u \bar{c} s$
decays~\cite{Gronau:1991dp}. In the case of charged $B$ decays, the
interference is between $B^-\to D K^-$ followed by a $D\to f$ decay and
$B^-\to \bar{D} K^-$ followed by $\bar{D}\to f$, where $f$ is any
final state common to both $D$ and $\bar{D}$. What makes this method
theoretically powerful is that there are no penguin contributions, and
all the hadronic unknowns are in principle obtainable from
experiment.

For our purpose, it is useful to group the different methods according
to the choice of the final state $f$, which can be (i) a CP-
eigenstate (e.g. $K_S \pi^0$, $K_S\phi$)~\cite{Gronau:1991dp}, (ii) a
flavor state ($K^+\pi^-$)~\cite{Atwood:1996ci}, or (iii) a multi-body
final state (e.g. $K_S\pi^+\pi^-$,
$\pi^+\pi^-\pi^0$)~\cite{Giri:2003ty}. Additional variations of the
basic method involve using multi-body $B$ decays (e.g. $B^+\to D
K^+\pi^0$)~\cite{Aleksan:2002mh}, use of $D^{0*}$ in addition to
$D^0$, self tagging $D^{0**}$ states~\cite{Sinha:2004ct} and neutral
$B$ decays (both time-dependent and
time-integrated)~\cite{Gronau:2004gt,Kayser:1999bu}.  Since these
measurements are statistically limited, an eventual combination of all
the modes will be needed in order to minimize the overall $\gamma$
measurement error~\cite{Soffer:1999dz, Atwood:2003jb}.
So far, the most precise direct
information on $\gamma$ comes from $B^\pm\to (K_S\pi^+\pi^-)_D K^\pm$,
where we use the notation $f_D$ to indicate a $D$ meson decaying into
the final state $f$. Both Belle~\cite{Poluektov:2004mf,Abe:2004gu} and
BaBar~\cite{Aubert:2004kv} have used $D^{*0}$ and $D^0$ decays, where
a subtlety of a sign difference between the two $D^{*0}$ decay modes
has been pointed out only recently~\cite{Bondar:2004bi}. These
measurements use a sum of Breit-Wigner resonances to model the Dalitz
plot distribution of $D^0\to K_S\pi^+\pi^-$. It is possible to remove
the associated modeling error by carrying out the measurements with a
model-independent treatment of the Dalitz
plot~\cite{Giri:2003ty,Atwood:2003mj}.

In all of the above methods, the Standard Model (SM) is
assumed. (Indeed, these methods involve only tree-level amplitudes
and, therefore, are unlikely to be affected by new physics.) Within
the SM, the largest theoretical uncertainty is due to $D-\bar D$
mixing. The parameters that describe the mixing are
\beq
x \equiv {\Delta m_D \over \Gamma_D} , \qquad 
y \equiv {\Delta \Gamma_D \over 2\Gamma_D}, \qquad
\theta \equiv \arg\left({q\over p}\right),
\eeq
where $\Delta\Gamma_D$ ($\Delta m_D$) is the decay width (mass)
difference between the two neutral $D$ mass eigenstates, $\Gamma_D$
($m_D$) is the average decay width (mass) of the mass eigenstates, and
$q$ and $p$ are the elements of the rotation matrix between the
interaction and mass eigenstate bases~\cite{Branco:1999fs}. (Here, and
in what follows, we choose a phase convention such that the tree-level
$D$ decay amplitudes are real, see Eq. \eqref{defrf}). The parameters
$x$ and $y$ cannot be calculated reliably in the SM. In particular,
one cannot rule out the possibility that they are as large as $x\sim
y\sim O(10^{-2})$~\cite{Dmix}, which is the range experiments are
beginning to probe~\cite{Aubert:2003pz,Burdman:2003rs}. A robust SM
prediction, however, is that $D-\bar D$ mixing is CP-conserving to a
very high accuracy, with a CP-violating phase of order $\theta\sim
O(10^{-4})$ \cite{Burdman:2003rs}.

If the $D-\bar D$ mixing parameters are known, their effect can be corrected
for in the measurement of
$\gamma$~\cite{Silva:1999bd,Meca:1998ee}. Without knowing their
values, assuming $x=y=0$ introduces an error in the extracted value of
$\gamma$. Making that assumption can introduce an error in the
determination of the branching ratios used in the ADS
method~\cite{Atwood:1996ci} of the order of $x/r_f\sim y/r_f \lesssim
20\%$, where $r_f$ is defined in Eq. (\ref{defrf}) below. Naively, one may
conclude that this introduces a similar error in the extracted value
of $\gamma$. It is the purpose of this note to explain why this is not
the case. We find that the effect is at most quadratic in $x$ and
$y$, thus it is very small and can be safely neglected.

Let us review the approach of extracting $\gamma$ neglecting $D-\bar
D$ mixing. We choose the phase convention in which
the $D$ meson decay amplitudes have the form
\beq \label{defrf}
A(D^0\to f)\equiv 
A_f,\qquad A(\bar D^0\to f)\equiv \bar A_{f}=A_f r_f
e^{-i\delta_f},
\eeq
such that $A_f$  and $r_f$ are positive.  
Since in the SM there is essentially no CP violation in
the $D$ decays, $\delta_f$ is a strong phase difference.  Due to the
abundance of flavor-tagged $D$ decays at the $B$ factories, the values
of $A_f$ and $r_f$ can measured very precisely from the decay rates
\beq\label{DGammaNoMix}
\Gamma_f\equiv\Gamma(D^0\to f)=A_f^2,\qquad 
\bar \Gamma_f\equiv\Gamma(\bar D^0\to f)= A_{f}^2 r_f^2.
\eeq
In the absence of $D-\bar D$ mixing, the amplitude for the
cascade decay $B^+\to f_D K^+$ is
\beq
A(B^+\to f_D K^+)=A_B A_f\left[r_f e^{-i\delta_f}+ r_B 
e^{i(\delta_B+\gamma)}\right],
\eeq
where $A_B$ and $r_B$ are positive parameters, $\delta_B$ is a strong
phase difference, and we have defined
\beq
A_B\equiv A(B^+\to \bar D^0 K^+),\qquad 
A_B r_B e^{i(\delta_B+\gamma)}\equiv A(B^+\to D^0 K^+).
\eeq
The sensitivity to $\gamma$ comes from the interference term in the
decay width
\beq\label{GammaBPl}
\Gamma(B^+\to f_D K^+)=A_B^2 A_f^2[r_f^2+ r_B^2+ 2 r_B r_f 
\cos(\delta_B+\gamma+\delta_f)].
\eeq
A similar expression for the $B^-\to \bar f_D K^-$ decay width is
obtained by making use of the absence of direct CP violation in the
$D$ decay:
\beq\label{GammaB}
\Gamma(B^-\to \bar f_D K^-)=
A_B^2 A_f^2[r_f^2+ r_B^2+ 2 r_B r_f \cos(\delta_B-\gamma+\delta_f)].
\eeq
Each final state $f$ introduces two new observables, 
$\Gamma(B^-\to \bar f_D K^-)$ and $\Gamma(B^+\to f_D K^+)$, 
and a single new unknown, $\delta_f$. 
The unknowns describing the $B\to D K$ part
of the cascade decay, $r_B$, $\delta_B$, and $ \gamma$, are the same
for all $D$ decay final states. Therefore, with enough $D$ decay modes,
there are more observables than unknowns, and the values of 
all the unknowns can be determined.

We now study the effect of CP-conserving $D-\bar D$
mixing. Specifically, we ask what error, $\Delta\gamma$, is introduced
in the extracted value of $\gamma$ when the analysis is done assuming
no $D-\bar D$ mixing.
A crucial ingredient in the approach described above for extracting
$\gamma$ is that $\delta_f$ is a pure phase, i.e., a single real
parameter. Specifically, we assumed that the absolute
magnitude of the interference term
\beq\label{interf}
\left|A_f \bar A_f^*\right|=\left|A_f^2 r_f e^{i\delta_f}\right|=A_f^2 r_f ,
\eeq
which is used in Eqs.\eqref{GammaBPl} and~\eqref{GammaB}, is already
measured in flavor-tagged $D$ decays. However, due to $D-\bar D$
mixing, time evolution dilutes the absolute magnitude of the
interference term
\eqref{interf}, which becomes another unknown. It is
the deviation of the magnitude of the interference term from its naive
value that introduces the error in the extracted value of $\gamma$.

In the presence of CP-conserving $D-\bar D$ mixing, the time-dependent 
$D$ decay amplitudes are \cite{Branco:1999fs}
\beq
\begin{split}
{\cal A}_f(t)={\cal A}(D^0(t)\to f)=g_+(t)A_f +g_-(t) \bar A_{f},\\
\bar{\cal A}_f(t)={\cal A}(\bar D^0(t)\to f)=g_+(t) \bar A_f +g_-(t)  A_f,
\end{split}
\eeq
where the time evolution functions are
\beqa
g_+(t)&=&\exp(-im_D t- \tau/2)\left[
\cosh\left({y\tau/ 2}\right)
\cos\left({x\tau/2}\right)+ i 
\sinh\left({y\tau/2}\right)
\sin\left({x\tau/2}\right)\right] \nonumber \\ &\approx& 
\exp(-im_D t-  \tau/2)\left[1+{(y+ i x)^2 \,\tau^2 / 4}\right],\nonumber \\
g_-(t)&=&\exp(-im_D t-  \tau/2)\left[
-\sinh\left({y\tau/ 2}\right)
\cos\left({x\tau/2}\right)- i 
\cosh\left({y\tau/2}\right)
\sin\left({x\tau/2}\right)\right] \nonumber \\ &\approx& 
\exp(-im_D t-  \tau/2)\left[{(-ix-y) \,\tau/ 2}\right],
\label{eq:evol}
\eeqa
with $\tau \equiv \Gamma_D t$. The approximations in
Eqs.~(\ref{eq:evol}) hold to second order in $x$ and $y$.
$D-\bar D$ mixing changes the time-integrated decay rates of
Eq. \eqref{DGammaNoMix} at leading order in $x$ and $y$. The precise change
is not important for our purpose. The key point is that the
time-integrated decay rates
\beq\label{AfRel}
\Gamma_f =\int dt |{\cal A}_f(t)|^2, \qquad 
\bar \Gamma_f=\int dt |\bar{\cal A}_{ f}(t)|^2,
\eeq
which are measured in tagged $D$ decays,
are exactly the rates that enter the $B$ decay
rate, which now reads
\beq\label{GammaBMix}
\Gamma(B^+\to f_D K^+)=A_B^2\left[\bar \Gamma_f+ r_B^2 \Gamma_f+ 
2 r_B \,\Re\!\left( e^{i(\delta_B+\gamma)} 
\int dt  {\cal A}_f(t) \bar{\cal A}_f(t)^*\right)\right].
\eeq
The impact of $D-\bar D$ mixing on  the $\gamma$ measurement occurs
only in the interference term 
\beq\label{relA}
\int d t {\cal A}_f(t)\bar {\cal A}_{f}(t)^*\equiv
\sqrt{\Gamma_f \bar \Gamma_f} e^{i\tilde \delta_f} e^{-\epsilon_f},
\eeq
where $\tilde \delta_f$ is a pure strong phase\footnote{Note that this is not
the case in the presence of CP-violating $D-\bar D$ mixing when the
phase $\tilde \delta_f$ is a combination of a weak and a strong
phase.} and
\beq \label{ImdeltafEq}
\epsilon_f= \frac{1}{8}(x^2+y^2) \left(\frac{1}{r_f^2}+r_f^2\right)-
\frac{1}{4}\big(x^2\cos 2\delta_f+y^2 \sin 2 \delta_f\big)
\eeq
describes the dilution due to $D-\bar D$ mixing.  The parameter
$\epsilon_f$ gives the approximate magnitude of the shift
$\Delta\gamma$ in the determination of $\gamma$. Since the leading
term in $\epsilon_f$ is proportional to $(x^2+y^2)/r_f^2$,
$\Delta\gamma$ is larger for cases where $r_f$ is smaller.  Apart from
the trivial case of no mixing ($x=y=0$), $\epsilon_f$ vanishes only if
$r_f = 1$ and either $y=0$ and $\delta = k\pi$, or $x=0$ and
$\delta=\pi/2 + k\pi$, where $k$ is an integer. In all other cases,
$\epsilon_f$ is positive.

In the special case $\epsilon_f=0$, there is no change in the $\gamma$
measurement. Each new mode $f$ still introduces only one new
parameter, $\tilde \delta_f$, which is obtained from the fit to the
decay widths. Moreover, the form of the equations is unchanged, as can be
seen by defining $\Gamma_f=\tilde A_f^2$, $\bar
\Gamma_f=\tilde A_f^2 \tilde r_f^2$ and 
comparing \eqref{AfRel}, \eqref{GammaBMix} with
\eqref{GammaBPl}. Therefore, for $\epsilon_f=0$, the correct value
of $\gamma$ is measured, even if the formalism used in the analysis ignores 
$D-\bar D$ mixing.

Our first main point is that while in general $\epsilon_f \ne 0$,
$\epsilon_f$ is of second order in the small parameters $x$ and
$y$. Therefore, its effect on the measurement of $\gamma$ is small.
Moreover, given measurements of, or upper limits on, $x$ and $y$, the
impact of $D-\bar D$ mixing can be accounted for without the need to
perform a time-dependent analysis of the $B$ decay.  This can be done
by using \eqref{ImdeltafEq} and \eqref{relA} in \eqref{GammaBMix}.

That $\epsilon_f$ is of second order in $x$ and $y$ can be understood
as follows. One can think of integration over time as a scalar product
in the vector space of time-dependent complex functions
\beq
\langle {\cal A}_f , \bar {\cal A}_f\rangle= \int
dt {\cal A}_f(t)\bar {\cal A}_{f}(t)^*\,.
\eeq
Then
\beq
|\langle {\cal A}_f , \bar {\cal A}_f\rangle|^2= \langle {\cal A}_f ,
{\cal A}_f\rangle \langle \bar {\cal A}_f , \bar {\cal A}_f\rangle
|\cos \Delta|^2,
\eeq
where $\Delta$ is a small angle linear in $x,y$. The
difference that defines $\epsilon_f$ in
Eq. \eqref{relA} is then
\beq
\epsilon_f \propto
\langle {\cal A}_f , {\cal A}_f\rangle \langle \bar {\cal A}_f ,
\bar {\cal A}_f\rangle - |\langle {\cal A}_f , 
\bar {\cal A}_f\rangle|^2\propto 1-\cos^2\Delta
 \sim O(\Delta^2) \sim O(x^2, y^2).
\eeq

We provide an explicit expression for $\Delta\gamma$ in one specific
case where $\gamma$ is extracted from a combination of a doubly
Cabibbo-suppressed decay width $\Gamma(B^\pm\to(K^\mp \pi^\pm)_D
K^\pm)$ and a decay width $\Gamma(B^\pm\to (f_{CP})_D K^\pm)$ into a
CP eigenstate.  To first order in $r_B$ and $r_f$ (here $f=K\pi$), we
get
\beq \label{shiftgamma}
\Delta\gamma=- \epsilon_f \times \frac{\cos \gamma \sin 2\gamma}{
\cos \gamma[\cos 2(\delta_f+\delta_B)- \cos 2 \delta_B]+
(r_B/r_f)\cos(\delta_f+\delta_B)[\cos 2\gamma-\cos 2\delta_B]}\,,
\eeq
where $\epsilon_f$ is taken from \eqref{ImdeltafEq}. For
$x^2+y^2\sim 2\%$, $\gamma\sim 60^\circ$, $r_B\sim 0.2$, and $r_f\sim
6\%$ we find the typical range
$\Delta\gamma\sim 0.1- 1^\circ$, depending on the values of the strong
phases $\delta_f$ and $\delta_B$. By the time the precision of the
$\gamma$ measurement reaches this level, we will have either
measurements or tighter upper limits on $x$ and $y$, so that the
measurement could be corrected for this shift.

Next, we consider the effect of $D-\bar D$ mixing in the case of
multi-body $D$ decays. For the Breit-Wigner treatment of Dalitz plot,
the corrections due to $D-\bar D$ mixing arise at $O(x^2, y^2)$ as in
the two-body case discussed above. Similar considerations apply in
both the two-body and three-body cases, with the difference being that
in the Breit-Wigner-based Dalitz plot analysis, $r_f$ varies over the
Dalitz plot.  The lowest value of $r_f$, and hence the largest
$\Delta\gamma$, is obtained in areas populated by doubly
Cabibbo-suppressed decays. Specifically, for the final state
$f=K_S\pi^+\pi^-$ this is the region of the decay $D^0\to
K^{*+}\pi^-$, which contributes most to the $\gamma$
measurement. Nonetheless, with $r_f$ of order a few percent in this
region, this still results in a small contribution to $\Delta\gamma$.
Moreover, the overall value of $\Delta\gamma$ is smaller due to the
contributions of other regions in the Dalitz plot in which $r_f$ is
larger.  The shift $\Delta\gamma$ is significantly smaller for singly
Cabibbo-suppressed multi-body decays, in which $r_f \sim {\cal
O}(1)$~\cite{Grossman:2002aq}.

Our second main point is that CP-conserving $D-\bar D$ mixing does not
affect the determination of $\gamma$ if the relevant Dalitz plot
parameters are determined by binning the Dalitz plot according to the
model-independent approach of Ref~\cite{Giri:2003ty}.  The phase space
integration over bin $i$ of the Dalitz plot introduces two new real
variables, $\hat c_i$ and $\hat s_i$:
\beq\label{relADalitz}
\hat c_i+i \hat s_i \equiv {c_i +  i s_i \over T_i}, 
\eeq
where
\beq
c_i+i  s_i \equiv \int_i dp \int d t 
{\cal A}_f(t)\bar {\cal A}_{f}(t)^*, 
\qquad 
T_i \equiv \int_i dp \int d t 
|{\cal A}_f(t)|^2.
\eeq 
The variables $c_i$ and $s_i$ are determined either from the binned
Dalitz plot obtained from the $B$ decay sample, or from
time-integrated decays of entangled $D$ states at a charm factory
operating at the $\Psi(3770)$~\cite{Giri:2003ty}. The point is that
measuring $c_i$ and $s_i$ already accounts for the dilution due to
$D-\bar D$ mixing. This is demonstrated by the fact that in the
two-body case, one can replace the two variables $\delta_f$ and
$\epsilon_f$ of \eqref{relA} with $\hat c_i$ and $\hat s_i$, which
satisfy $\hat c_i^2+\hat s_i^2=1 - {\cal O}(x^2, y^2)$. The method of
Ref.~\cite{Giri:2003ty} is already designed to handle $\hat c_i^2+\hat
s_i^2 < 1$, which in multi-body decays arises due to the phase space
integration over each bin.

We concentrated on the case of CP-conserving $D-\bar D$ mixing, since
this is the case in the SM. With new physics, this may not be the
case. Then, our results do not hold and larger effects are
introduced. For example, consider the case where there is new physics
in the mixing, with a CP-violating phase $\theta \sim O(1)$.  
Then the assumption
of no $D-\bar D$ mixing introduces an error in the value of $\gamma$
of order $\Delta\gamma \sim O(x\,\theta,y\,\theta)$ which is
linear in the small parameters.

It is instructive to compare our results to those of
\cite{Silva:1999bd}. The analysis in \cite{Silva:1999bd} corresponds
to a situation in which the $D$ decay amplitudes were determined from
$D$ decay data by taking $D-\bar D$ mixing into account, but
neglecting mixing in the $B$ decay analysis. The error in the value of
$\gamma$ extracted in this way is linear in $x$ and $y$, regardless of
whether $D-\bar D$ mixing is CP-conserving or not.  Here, on the other
hand, we show that when both the $D$ and the $B$ decay amplitudes are
extracted ignoring $D-\bar D$ mixing, CP conserving $D-\bar D$ mixing
induces an error in the extracted value of $\gamma$ that is of second
order in $x$ and $y$. This provides a simpler practical approach for
solving the problem introduced by CP-conserving $D-\bar D$ mixing,
which is the case in the SM.

To conclude, we show that within the SM, neglecting $D-\bar D$ mixing
in the extraction of $\gamma$ using $B\to D K$ type decays introduces
at most an $O(x^2,y^2)$ effect. This is a very small effect that can be
neglected for all practical purposes.

\bigskip\medskip\noindent
{\bf ACKNOWLEDGMENTS}

\bigskip
We thank Jo\~{a}o Silva for helpful discussions and comments.  Y.G. and
J.Z thank the Institute of Nuclear Theory at the University of
Washington for its hospitality and the Department of Energy for
partial support during the completion of this work. The work of
J.Z. is supported in part by the Department of Energy under Grants
DOE-ER-40682-143 and DEAC02-6CH03000.  The work of A.S. is supported
by the U.S. Department of Energy under contract DE-FG03-93ER40788.

\end{document}